\documentclass[useAMS,usenatbib]{mn2e}
\usepackage{natbib}
\usepackage{graphicx}
\usepackage{color}
\usepackage{enumerate}
\usepackage{verbatim}  % for using the block comment command
\usepackage{booktabs}
\usepackage{amsmath}
\usepackage[table]{xcolor}
\usepackage[section]{placeins}
\usepackage{float}
\usepackage{bm}
\begin{document}
\title[Limits on single-source GWs from B1937+21]{Limits on the strength of individual gravitational wave sources using high-cadence observations of PSR B1937+21.}
\author[S. X. Yi et al.]{Shuxu Yi$^{1,2}$, Benjamin W. Stappers$^{3}$,
Sotirios A. Sanidas$^{3}$,
Cees G. Bassa$^{3,4}$,\and
Gemma H. Janssen$^{3,4}$,
Andrew G. Lyne$^{3}$,
Michael Kramer$^{5}$,
Shuang-Nan Zhang$^{1,6,7}\thanks{E-mail:
zhangsn@ihep.ac.cn}$\\
$^{1}$ Key Laboratory of Particle Astrophysics, Institute of High Energy Physics, Chinese Academy of Sciences, Beijing 100049, China\\
$^{2}$ University of Chinese Academy of Sciences, Beijing 100049, China.\\
$^{3}$ Jodrell Bank Centre for Astrophysics, University of Manchester, Manchester, M13 9PL, United Kingdom\\
$^{4}$ ASTRON, the Netherlands Institute for Radio Astronomy,
Postbus 2, 7990 AA Dwingeloo, the Netherlands\\
$^{5}$ MPI f\"ur Radioastronomie, Auf dem H\"ugel 69, 53121 Bonn, Germany\\
$^{6}$ Space Science Division, National Astronomical Observatories of
China, Chinese Academy of Sciences, Beijing 100012, China\\
$^{7}$ Physics Department, University of Alabama in
Huntsville, Huntsville, AL 35899, USA\\}

\maketitle
\begin{abstract}
We present the results of a search for gravitational waves (GWs) from individual sources using high cadence observations of PSR B1937+21. The data were acquired from an intensive observation campaign with the Lovell telescope at Jodrell Bank, between June 2011 and May 2013. The almost daily cadence achieved, allowed us to be sensitive to GWs with frequencies up to $4.98\times10^{-6}\,\rm {Hz}$, extending the upper bound of the typical frequency range probed by Pulsar Timing Arrays. We used observations taken at three different radio frequencies with the Westerbork Synthesis Radio Telescope in order to correct for dispersion measure effects and scattering variances. The corrected timing residuals exhibited an unmodeled periodic noise with an amplitude $~150\,\rm {ns}$ and a frequency of $3.4\rm {yr}^{-1}$. As the signal is not present in the entire data set, we attributed it to the rotational behaviour of the pulsar, ruling out the possibilities of being either due to a GW or an asteroid as the cause. After removing this noise component, we placed limits on the GW strain of individual sources equaling to $h_{\rm s}=1.53\times10^{-11}$ and $h_{\rm s}=4.99\times10^{-14}$ at $10^{-7}\,\rm {Hz}$ for random and optimal sources locations respectively.
\end{abstract}


\section{Introduction}
High-precision timing of millisecond pulsars has long been proposed as one of the methods to directly detect gravitational waves (GWs hereafter) \citep{Sazhin,Detweiler,HD,J05}. Pulsar timing can detect GWs in the frequency range from approximately $T_{\rm{obs}}^{-1}$ to $N(2T_{\rm{obs}})^{-1}$, where $T_{\rm{obs}}$ is the time span of the observations and $N$ is the number of observations. Typically, this frequency window extends from $\sim10^{-7}\,{\rm Hz}$, corresponding to a bi-weekly observation scheme, to $\sim10^{-9}\,{\rm Hz}$, corresponding to a $T_{\rm{obs}}$ of several years \citep{Haasteren11,Manchester13}. The most prominent GW sources in this frequency range are coalescing super-massive black hole binaries (SMBHBs) located in the nuclei of merged galaxies. As there are expected to be a large number of such systems, their signals overlap and form a stochastic GW background (SGWB) \citep{Phinney01,JaffeBacker,WyitheLoeb}. Considerable effort has been made to detect or limit the SGWB using single pulsars or a pulsar timing array (PTA) \citep{Kaspi94,mchugh96,lommen02,jenet06,Haasteren11,demorest13,ShannonSci}, placing more and more stringent upper limits on its amplitude. Recently \citep{Sesana13}, it has been argued that the most up-to-date SGWB limits from PTAs are close to the theoretically expected values of the SMBHB SGWB, making its imminent detection possible.

\begin{table*}
\centering
\caption{Summary of observations used in this analysis.}\label{obs}
\scalebox{0.9}{
\begin{tabular}{|c|c|c|c|c|c|}
\hline
 dataset&WSRT350&WSRT1380&WSRT2273&LT&42FT\\
\hline

Telescope&\multicolumn{3}{c|}{Westerbork Synthesis Radio Telescope}&Lovell&42-foot @ Jodrell Bank\\

\hline
Backend&\multicolumn{1}{r}{}&\multicolumn{1}{c}{PuMaII}&&ROACH&COBRA2\\
\hline
Frequency (MHz)&350&1380&2273&1532&610\\
\hline
Bandwidth(MHz)&80&160&160&400&5 or 10\\
\hline
Data span&5/2011 - 5/2013&5/2011 - 4/2013&5/2011 - 4/2013&11/2011 - 4/2013&6/2011 - 5/2013\\
\hline
Number of observations&26&31&21&450&509\\
\hline
Averaged ToA uncertainty (s)&$3.37\times10^{-7}$&$4.91\times10^{-8}$&$2.95\times10^{-7}$&$8.23\times10^{-8}$&$1.86\times10^{-6}$\\
\hline
\end{tabular}}
\end{table*}

Modeling of the expected GW background shows that GWs from close and/or massive enough SMBHB systems will stand out from the background as resolvable signals \citep{Sesana09}. An increasing number of investigations has been performed in the recent years concerning the expected amplitude of individual GW sources: \cite{lommen&backer01} searched for possible GW induced timing residual variations in PSRs B1937+21 and J1713+0747. The GWs they were searching for were from a presumed SMBHB at Sagittarius A*, which might have been responsible for the $\sim$106 day quasi-periodic radio flux variations \citep{Zhao01}. They found no timing residuals larger than 150\,ns at the corresponding period of 53 days in their data, which means the strain of GWs emitted from Sgr A* should be less than $1.37\times10^{-13}$ at a frequency of $2.18\times10^{-7}\,{\rm Hz}$, assuming an optimal polarization. \cite{J04} derived general expressions for the expected timing residuals induced by GWs emitted from a slowly evolving SMBHB, and ruled out the proposed SMBHB system in 3C 66B \citep{Sudou} with 95\% confidence since no expected fluctuations were observed in the timing residuals of PSR J1857+0943. \cite{Yardley10} (Y10 hereafter) determined the sensitivity of the Parkes PTA to GWs emitted by a non-evolving individual SMBHB from $10^{-9}\,{\rm Hz}$ to $4\times10^{-7}\,{\rm Hz}$ using the timing results of 18 pulsars from the \cite{verbiest09} dataset. The data were most sensitive to GWs at the frequency of $\sim9\times10^{-9}\,{\rm Hz}$, with GW strain larger than $\sim9\times10^{-14}$.

Detecting GWs from individual sources can provide useful astronomical information, such as the sky position, the chirp mass-distance combination of the sources \citep{Sesana10} and limit the merger rate of nearby SMBHBs \citep{wen11}. Furthermore, the pulsar timing parallax signal enables us to take advantage of the pulsar term in the timing residuals to determine more parameters such as the distance, the chirp mass and the spin of the SMBHB \citep{Corbin10,Chiara12} separately, and increase the accuracy of pulsar distance measurements \citep{LKJ11}.

The likelihood that more individual sources are resolvable at frequencies above $10^{-8}\,{\rm Hz}$ \citep{Sesana08} argues strongly for performing observations of pulsars at a higher cadence. The Monte Carlo simulation of \cite{Sesana09} predicts the strain of the GWs from resolvable single sources to be $\sim10^{-15}$. Here we present our dedicated campaign using PSR B1937+21. This pulsar was chosen despite the fact that it is known to exhibit long term timing noise and time-dependent dispersion measure (DM) variations \citep{Cordes90,Kaspi94,Ramachandran} because it is bright, isolated and relatively consistent in flux. Additionally, the effects of DM variations are potentially correctable. We use 528 pulse times-of-arrival (ToAs) over the span of 650 days to obtain the timing solution, resulting in an average time between ToAs of just 1.23 days, and more typically the pulsar was observed daily or twice daily. With such high cadence observations, we managed to extend our sensitivity to GWs with frequency up to $4.98\times10^{-6}\,{\rm Hz}$, which is in the frequency regime so far probed only by observations of close binary systems \citep{hui12}.

The paper is organized as follows. The details of the observations are presented in Section \ref{observations}. In Section \ref{fitting}, we describe the procedure for correcting for the interstellar medium (ISM) effects, such as DM and scattering variations,  and the modeling of the ToAs. Then we use the resulting timing residuals to evaluate the sensitivity to individual GW sources in Section \ref{result}. In the discussion section, we elaborate on the results and on the possible origins of the unmodeled components which were observed in the ToAs.

\section{Observations}

\begin{figure}
\centering
\includegraphics[width=9 cm]{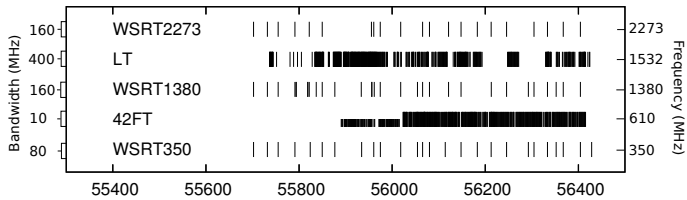}
\caption{Schematic plot of the dates on which ToAs were obtained. See text for details. }\label{cadence}
\end{figure}

The details of the observations are summarized in Table \ref{obs}. In our analysis, five datasets from three telescopes were used. The first two datasets are from Jodrell Bank Observatory, obtained with the 76\,m Lovell telescope (LT, hereafter) and  the 42\,ft (13\,m diameter) telescope (42FT), with the ROACH and COBRA2 backends, both performing online folding and coherent dedispersion using \texttt{DSPSR} \citep{DSPSR}. The ROACH backend of the Lovell telescope observes a $\sim$400\,MHz wide band at L-band, centered at 1532\,MHz \citep{Karuppusamy}. The COBRA2 backend of the 42\,ft telescope observes at a center frequency of 610\,MHz, initially with a bandwidth of 5\,MHz but this was doubled to 10\,MHz from the beginning of April 2012. The time span of the LT dataset is from June 2011 to May 2013, consisting of 450 observations. The 42FT dataset has 509 observations from November 2011 to April 2013. In order to improve the signal-to-noise ratio (SNR), the daily ToAs are combined in groups of eight using the binning tools provided by \texttt{TEMPO2} \citep{hobbs06}, 63 ToAs are produced from the 42FT dataset. The three other datasets are from the Westerbork Synthesis Radio Telescope (WSRT) using the PuMaII backend \citep{Karuppusamy08}. Observations were taken approximately monthly at three frequencies, centered at 350\,MHz, 1380\,MHz and 2273\,MHz respectively. The bandwidths used for these frequencies are either 80\,MHz or 160\,MHz, and are listed in Table \ref{obs}. For this study we have used observations taken in the time span from May 2011 till May 2013.
The observations were stored offline and coherently dedispersed and folded using DSPSR, and subsequent manipulation was done using \texttt{PSRCHIVE}. For each observation multiple data products were stored with varying frequency and time resolution. To generate ToAs, fully scrunched profiles were cross-correlated with an analytic template created by \texttt{PSRCHIVE} based on integrated high-signal-to-noise observations at each frequency. For all telescopes and frequencies the ToAs were referred to local time using a Hydrogen maser at the observatory, then converted to TT(TAI) using the GPS system and corrections on UTC as provided by the \texttt{BIPM}\footnote{http://www.bipm.org}. A schematic plot of the dates on which ToAs were obtained are plotted in Fig.~\ref{cadence}. We have used only the total intensity data for this study by summing the power measured in two feeds.

%s\FloatBarrier
\label{observations}
%\FloatBarrier
\section{Data processing}
\label{fitting}

\subsection{Correction for DM variations and pulse profile broadening}
\begin{figure}%%[h]
\centering
\includegraphics[width=8 cm]{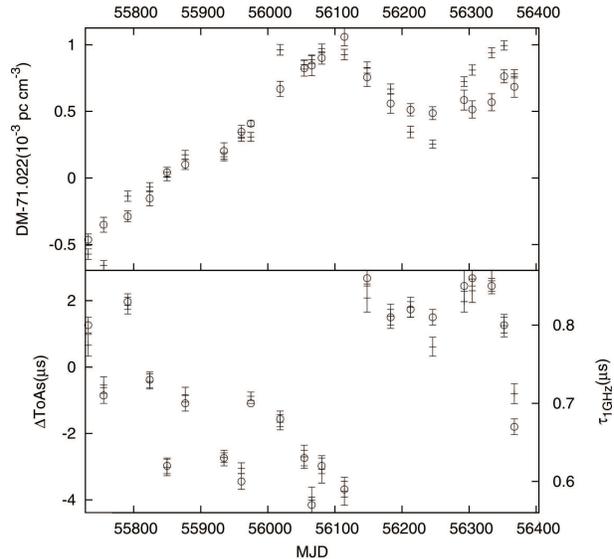}
\caption{Upper panel: DM as a function of MJD; a constant DM of $71.022$ is subtracted for convenience of plotting. The plus marks with error bars are DM taken from fitting the 512 frequency channel WSRT350 data, the circles are from fitting DMOFF (see text) with the three datasets with the lower frequency (WSRT350, 42FT, WSRT1380). Bottom panel: The difference between ToAs obtained with a constant template and the ToAs that take time-dependent scattering into consideration (crosses). The $\tau_{1{\rm GHz}}$ as a function of date is also plotted (circles).}\label{dmcompare}
\end{figure}

PSR B1937+21 shows significant DM variations, which are the main source of timing noise at the level of microseconds at the frequency of $10^3\,\rm{MHz}$ over a time span of a couple of years. To attempt to remove this noise, we need to evaluate the DM value at each observation epoch. For this purpose, we chose to use the WSRT350 dataset because it has the best combination of SNR and fractional frequency difference for determining the DM. Here we will describe the process we followed: First, we started by using all 512 frequency channels of the WSRT350 observations. Then we selected one observation, aligned it in pulse phase across all frequency channels and collapsed them in one channel to make a pulse profile template. The template was cross-correlated with the observed profiles across all frequency channels, generating 512 ToAs for each epoch. These ToAs were fitted with TEMPO2 for DM, and we obtained a DM($t_i$) value for every observation $i$. The DM($t_{\rm i}$) as a function of MJD is plotted in the upper panel of Fig.~\ref{dmcompare}.

\begin{figure}%[h]
\centering
\includegraphics[width=7 cm]{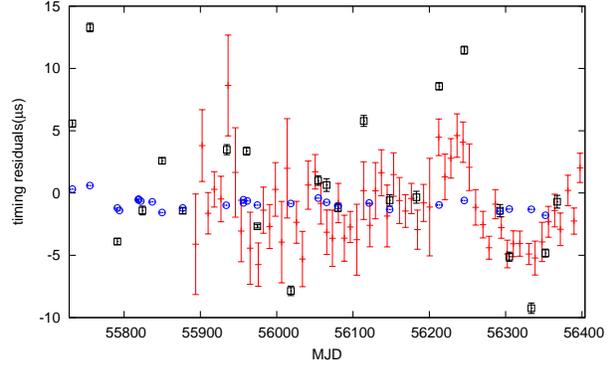}
\caption{Timing residuals of WSRT1380 (blue circles), 42FT (red dots) and WRST350 (black squares) after the first step of correction of DM measured from the WSRT350 data. The frequency dependent structure comes from offsets of the DM($t_{\rm i}$) (see text for details).}\label{residmofff350}
\end{figure}

These DM($t_{\rm i}$) values were then used to re-dedisperse the WSRT350 profiles before they were collapsed to form a single profile per epoch. Using these improved profiles, we formed a new template and cross-correlated that with each individual observation to generate new ToAs. Comparing these ToAs with the timing solution including the DM($t_i$) at each epoch, we created a set of residuals that retain some unmodeled structure (See Fig. \ref{residmofff350}). This implies that the DM correction using WRST350 along does not sufficiently remove the effect of ISM; therefore further investigations are necessary.

Scattering in the ISM may cause the pulse profiles to be broadened and this effect is more pronounced at low frequencies. While the scattering will affect the shape of the pulse profile, if it remains constant, and is accounted for in the template used to generate the ToAs, the timing should be unaffected. However, if the degree of scattering changes then this could lead to timing noise.

\begin{figure}
\centering
\includegraphics[width=8cm]{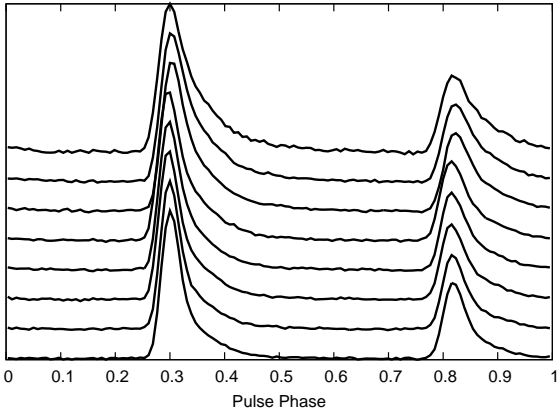}
\caption{Pulse profiles of eight frequency channels on MJD 56428, with rising edges aligned. From bottom to top, the center frequencies (MHz) are: 376.287, 367.500, 358.749, 349.998, 341.250, 332.498, 323.748, {\rm and} 314.960. The tails of the profiles exhibit increasing broadening with decreasing frequencies.}\label{scatteredprofile}
\end{figure}

\begin{figure}%[h]
\centering
\includegraphics[width=8cm]{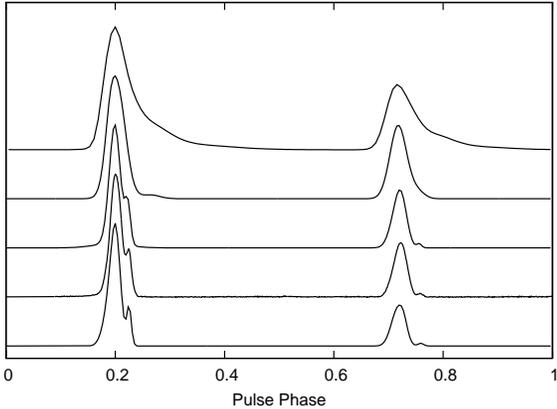}
\caption{Template profiles for different datasets, from bottom to top: WSRT2273, LT, WSRT1380, 42FT, WSRT350. The profiles have arbitrary offsets with respect to each other. The main pulses have been normalized to the same height.}\label{template}
\end{figure}

To determine the influence of scattering on our DM determinations and ToAs we adopted the following procedure: First, we collapsed the 512 frequency channels of the WSRT350 data into 8 frequency bands. An example of the eight profiles, obtained on MJD 56428, is shown in Fig.~\ref{scatteredprofile}. Then we constructed a frequency dependent template by convolving a reference profile with a truncated exponential, i.e.,
$\exp(-\phi P/\tau_{\nu})$, where $\phi$ is the pulse phase, $P$ is the period of the pulsar and $\tau_{\nu}$ is the broadening timescale at frequency $\nu$. The broadening timescale is related with frequency by
\begin{equation}
\frac{\tau_{\rm \nu}}{\tau_{1{\rm GHz}}}=\left(\frac{\nu}{1{\rm GHz}}\right)^{-4},
\end{equation}
where $\tau_{1{\rm GHz}}$ is the value of $\tau_{\rm \nu}$ at $\nu=1{\rm GHz}$. For Kolmogorov thin-screen media, the scattering time is expected to be proportional to $\nu^{-4.4}$. However, the measured index is closer to -4 than -4.4, which can be find in section 4.2.1 of \cite{handbook} and \cite{Ramachandran}. The reference profile we used is the profile of the 42FT data at 610\,MHz. The reason not to use the profile at higher frequencies is that the amplitude ratio between the main pulse and interpulse evolves significantly with frequency, as is shown in Fig.~\ref{template}. The use of the profile of the 42FT data at $610\,\rm{MHz}$ as the reference profile provides the best compromise between reducing the effect of scattering and the least profile evolution. We then search for the value of $\tau_{1{\rm GHz}}$ that makes the convolved templates best fit with the observed profiles and thus get a time-dependent pulse broadening timescale. The broadening timescale at  $1\,GHz$ as a function of time is plotted in the bottom panels of Fig.~\ref{dmcompare} as circles. The ToA of each observation is then obtained, by using the convolution of the standard profile and the truncated exponential $\exp(-\phi P/\tau_{\nu\rm{c}})$ as a template, where $\tau_{\nu\rm{c}}$ is the broadening timescale of the central frequency of the band. The difference between ToAs measured without and with the removal of the scattering effect, $\Delta\rm{ToAs}$, is plotted in the bottom panel of Fig.~\ref{dmcompare}. A very clear correlation exists between $\Delta\rm{ToAs}$ and $\tau_{1{\rm GHz}}$. The physical origin of this correlation is straightforward: when cross-correlating a constant profile template with a pulse profile that has a broader tail, the resulted ToA is overestimated and vice versa. The difference in ToAs can be as large as $4\,{\rm \mu s}$, which highlights the importance of correcting for scattering.

The scattering effects become less significant at higher frequencies, as $\tau_{\nu}\propto f^{-4}$. Therefore, for these datasets the ToAs are measured following the usual procedure using their own templates without correcting for scattering. The template profiles are obtained by adding together pulse profiles at each epoch and fitting to analytical models with \texttt{paas} \citep{PSRCHIVE2}. The template profiles for each dataset are shown in Fig.~\ref{template}.

The ToA residuals of the lowest frequency bands after correcting for the DM variations are plotted in Fig.~\ref{residmofff350}. The timing residuals of three frequency bands show evidently a common structure, and the amplitude of the structure decreases as frequency increases. This suggests that the DM$(t_{\rm i})$ we are using have some offsets, $\Delta\rm{DM}(t_{\rm i})$, from the real values. Therefore, we fit for a list of $\Delta\rm{DM}(t_{\rm i})$ using ToAs of these three bands.  The fitting process is to add offsets on DM($t_{\rm i}$) into the timing model, so as to remove a common trend among the residuals of all bands \citep{keith13}, i.e.,
\begin{equation}
\Delta t_{\rm i}\approx 4.15\times10^6\times f^{-2}\times \Delta\rm{DM}(t_{\rm i}).\label{dmtoa}
\end{equation}
We do not measure $DM(t_i)$ and the scattering timescale simultaneously with the WSRT350 data, because that results in an insufficiently precise determination of the DM, as the frequency range is not large enough to disentangle the two effects. The new DM($t_{\rm i}$) values with $\Delta\rm{DM}(t_{\rm i})$ added are plotted in the upper panel of Fig.~\ref{dmcompare} as circles. The DM offsets might also arise from ISM scattering effects, However, we failed to find a significant correlation between $\Delta\rm{DM}(t_{\rm i})$ and $\tau_{1{\rm GHz}}$ to support that. %In the bottom panel of Fig.~\ref{dmcompare} we plot $\Delta\rm{DM}$ along with the $\tau_{1{\rm GHz}}$ as a function of date. %As shown in the plot, the variations of $\Delta\rm{DM}$ show a weak correlation with $\tau_{1{\rm GHz}}$. The Pearson product-moment correlation coefficient, i.e, a measure of the linear dependency between $\Delta\rm{DM}$ and $\tau_{1{\rm GHz}}$ is 0.489, proving the correlation \citep{Pearson89}. We explain the physical origin of this correlation as follows: the frequency-dependent profile broadening timescale results in an additional frequency dependency of the ToAs, which contributes to $\Delta\rm{DM}$.

With DM offsets corrected, the timing residuals of WSRT2273, LT, WSRT1380 and WSRT350 are plotted in the upper panel of Fig.~\ref{comparesi}. All residuals track well with each other, which means there is no significant DM error remaining.

\subsection{Timing model}
\begin{figure}
\centering
\includegraphics[width=\columnwidth]{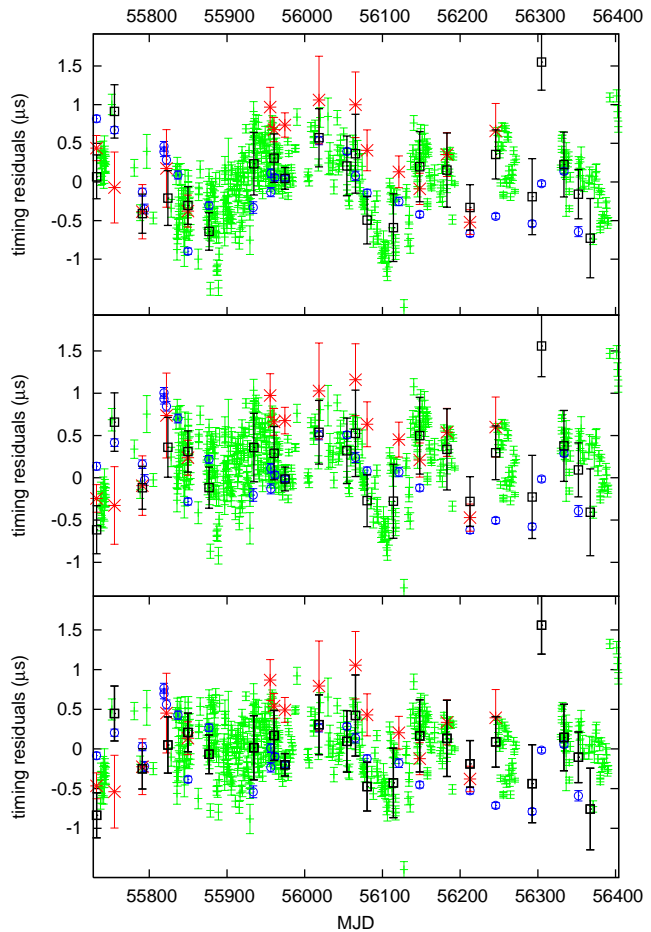}
\caption{Upper panel: Timing residuals of WSRT2273 (red stars), LT (green dots), WSRT1380 (blue circles) and WSRT350 (black squares) after DMOFF correction. Middle panel: The same plot with the timing model parameters listed in Table \ref{para}, using the same symbols as the upper panel. Bottom panel: the same plot as above, after wave components removed. The parameters of the wave harmonics are listed in Table \ref{waves}}\label{comparesi}
\end{figure}

\begin{table}%[h]
\centering
\caption{The best-fit parameters of PSR B1937+21.}\label{para}
\begin{tabular}{l l}
\hline
Parameters & Values\\
\hline
Reference Epoch & 55965\\
$\alpha\rm{(J2000)}$(hms) & 19:39:38.561336(1)\\
$\delta\rm{(J2000)}$(dms) & 21:34:59.12596(1)\\
$\nu$ ($\rm{s}^{-1}$) & 641.928220971573(2)\\
$\dot{\nu}$ ($10^{-14}\times\rm{s}^{-2}$) & -4.33103(2)\\
$\mu_{\alpha}$ (mas $\rm{yr}^{-1}$) & -0.15(3)\\
$\mu_{\delta}$ (mas $\rm{yr}^{-1}$) & -0.18(5)\\
\hline
\end{tabular}
\end{table}
After the correction of DM variations and pulse profile broadening, we use ToAs from all datasets except the 42FT to fit for the pulsar coordinates (RAJ, DECJ), rotation frequency (F0), first time derivative (F1) and proper motion (PMRA, PMDEC) with other parameters fixed. The reason to exclude the 42FT data is that it has the lowest SNR, and the uncertainty of the ToAs are an order of magnitude larger than in the  other datasets. The best-fit parameters are listed in Table \ref{para}. Parameters in Table \ref{para} are broadly consistent with values in \cite{verbiest09}. However, as our time-span is only 2 years, there are strong correlations between the proper motion and the position and the pulsar spin down, in our case $\nu$, $\dot{\nu}$. Therefore the errors of the parameters are underestimated. The timing residuals of all datasets are plotted in the middle panel of Fig.~\ref{comparesi}.

\begin{table}
\centering
\caption{Harmonic wave parameters. The wave epoch is set at MJD=55965}\label{waves}

\scalebox{1}{%
\begin{tabular}{llll}
\toprule
 & WAVE1 & WAVE2 & WAVE3\\
\midrule
frequency ($\rm{yr}^{-1}$) & 3.4 & 6.8 & 10.2 \\
Amplitude (ns) & 150(20) & 58(29) & 105(27) \\
Initial phase (rad) & 0.19(12)& 1.2(3)& 0.0(1)\\
%sin amp (ns) & 152.38 & -18.73 & -105.14 \\
\bottomrule
\end{tabular}}
\label{table2}
\end{table}

The root mean square (RMS) post-fit residual is 0.346 $\mu\rm{s}$. The timing residuals exhibit trends of unmodeled periodic signals with amplitude around $150\,\rm{ns}$, which unveil themselves in the power spectrum as distinct peaks at low frequencies (see the upper panel line in Fig.~\ref{power}). Their effect is further exacerbated at low frequencies by the spectral leakage that Lomb Scargle periodograms can suffer from. The highest peak occurs at the frequency of $3.4\,\rm{yr}^{-1}$ $(1.078\times10^{-7}\,\rm{Hz})$. The power spectrum of the timing residuals is generated using a Lomb-Scargle periodogram \citep{press&rybicki}, which was designed for handling unevenly sampled time sequence data.

The peaks with high significance at low frequencies can be removed by applying harmonic waves fitting \citep{Hobbs04} up to the third harmonic. The fundamental frequency is $3.4\,\rm{yr}^{-1}$ $(1.078\times10^{-7}\,\rm{Hz})$. We set the wave epoch to be MJD=55965. The results of wave-fitting are listed in Table \ref{table2}.
The timing residuals after harmonic wave fitting is plotted in the bottom panel of Fig.~\ref{comparesi}, and their Lomb-Scargle periodogram is plotted with in the lower panel of Fig.~\ref{power}. \\

\section{Sensitivity to individual GW sources}\label{result}
In this section we evaluate the sensitivity of the timing residuals to individual GW sources. We assume that there are either none or negligible GW signals in our timing residuals. This assumption is also strengthened by a whitening process  which will be described in Section 4.1. Then, we use these timing residuals in order to create a noise template for the pulsar. The sensitivity curve is then produced by injecting monochromatic GWs in our initial dataset and running a detection algorithm. The detection algorithm is executed repetitively by increasing the strain of the induced GW signal until a detection is made. The detection is determined by the noise template and a detection threshold, which is dictated by the false alarm probability we have chosen.
\subsection{The noise model and the signal detection threshold}

We follow the same process as in Y10, which we briefly recap here. First, we take the logarithm of the power in the power spectrum of the corrected timing residuals. Then, we smooth the log-power spectrum using a moving average technique. The noise template is then provided by a fourth-order polynomial which is used to represent the smoothed power spectrum. With this process, we ensure that we have removed any spurious effects from unmodeled signals in the timing residuals.
\newline
\begin{figure}
\centering
\includegraphics[width=8 cm]{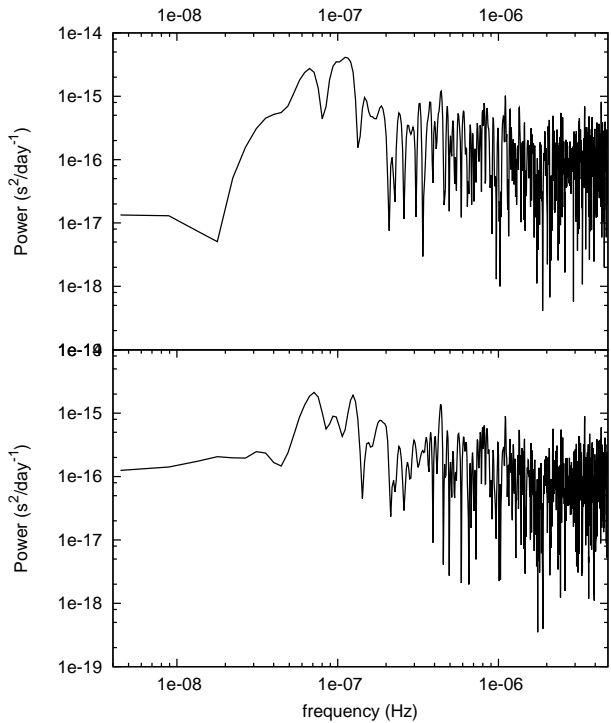}
\caption{Power spectra of the post-fit timing residuals. The ordinates are normalized such that the total power is equal to the variance of the timing residuals. The upper panel is the power spectrum calculated before the harmonic wave removal, corresponding to the middle panel of Fig.~\ref{comparesi}. The lower panel is after the removal of the harmonic waves, corresponding to the bottom panel of Fig.~\ref{comparesi}. The highest peak in the upper curve is at the frequency of $3.4\,\rm{yr}^{-1}$ $(1.078\times10^{-7}\,\rm{Hz})$.}
\label{power}
\end{figure}

The signal detection threshold is set as the product of the noise template and a scaling factor, below which any peak is considered to be due to noise. The scaling factor is denoted as $\alpha$, which is determined by simulation as follows. $10^4$ ToA datasets with white noise of $100\,\rm{ns}$ RMS and the same sampling as the real data were generated. Then we computed the respective power spectra and for each one of them we calculated the mean power. We label the mean power of the $i$th power spectrum with $m_{\rm i}$. Then, we assigned increasing values to a scaling factor $\alpha$ starting from $\alpha=1$; for each trial of $\alpha$, we count the number of power spectra which has any power value higher than $\alpha m_{\rm i}$. We adjust the value of $\alpha$ until the percentage of such power spectra of all realizations is equal to the established false-alarm rate. $\beta=\rm{ln}\alpha$ is added to the fourth-order polynomial, i.e, the noise template in log scale, to form a threshold over the whole frequency range. We found $\alpha=10.8$, when each detection of single GWs (without the target frequency known in advance) has a false-alarm rate of one percent.

\subsection{Sensitivity curve}
\begin{figure}
\centering
\includegraphics[width=8.5 cm]{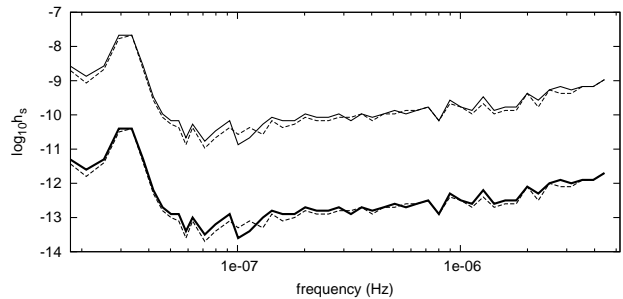}
\caption{The sensitivity to single GW sources in the observations of PSR B1937+21. The upper two lines are sensitivity to GW sources of random sky location and polarization with (thin dashed line) and without (thin solid line) harmonic waves fitting in the timing model. The bottom two lines are sensitivity to GW sources of optimal sky location and polarization with (thick dashed line) and without (thick solid line) the harmonic waves (Table~\ref{table2}) fitted in the timing model. The finite sensitivity at the frequency of $1\,{\rm yr}^{-1}$ is due to the limited resolution of our sample on the curves.}
\label{curve}
\end{figure}

The process followed to construct the GW sensitivity curve was presented in detail in Y10. In short, we inject a GW signal in the TOAs and then take the post-fit timing residuals. Then we compare the power spectrum of the timing residuals with the detection threshold and we repeat this process continuously increasing the strain of the injected GW until we have a detection. In the rest of this section we describe the whole process in detail. The frequency range between $1/T_{\rm{obs}}$ and $N(2T_{\rm{obs}})^{-1}$ is evenly divided into 50 bins in log scale. The detection algorithm that we describe here is applied in each one of these frequencies respectively.

First we add a sinusoid into the ToAs with an amplitude
\begin{equation}
A=\frac{h_{\rm s}}{\omega}(1+\cos\theta)\sin(2\phi)\sin\left[\frac{\omega D_p(1-\cos\theta)}{2c}\right],
\label{amplitude}
\end{equation}
where $h_{\rm s}$ is the strain of the injected GW, $\theta$ is the angle between the GW source and the pulsar, $\phi$ is the GW polarization angle, $\omega=2\pi f$ is the angular frequency of GWs and $D_{\rm p}$ is the distance from the Earth to the pulsar. The start value for the strain is $10^{-15}$ and we increase it with a step $\Delta \log h_{\rm s}=0.1$ until we make a detection. The values of $\theta$ and $\phi$ are assigned randomly at each step so that the probability density of the source position is uniform on the celestial sphere. For typical values of $D_{\rm p}$ (several kpc) and $\omega$ (several $\rm{year}^{-1}$), the amplitude $A$ oscillates dramatically with $\theta$ according to equation~\eqref{amplitude}. Therefore, the optimal position of a GW source at which $A$ takes its greatest value is somewhere very close to the location where $\theta=0$ and $\phi=\pi/4$. The $D_p$ we used is $3.55\,{\rm kpc}$, which is estimated using the average value of $DM$ through
\begin{equation}
DM=\int_0^{D_{\rm p}}n_{\rm e}ds,
\end{equation}	
where $n_{\rm e}$ is the density distribution of free electrons along the light of sight taken from the NE2001 model \citep{NE2001}\footnote{The web interface to do this evaluation: \newline http://www.nrl.navy.mil/rsd/RORF/ne2001/\#los}.

Then we use TEMPO2 to get the post-fit residuals from the GW-injected ToAs. In the fitting process, we let RA, DEC, F0, and F1 to be free parameters and keep all the other parameters fixed. Since the influence of the simulated GWs are not frequency dependent, further fitting for DM is not needed. The power spectrum of the residuals was calculated using a Lomb-Scargle periodogram in the next step. If the power in the neighborhood of the current frequency bin surpasses the detection threshold that we have set, then a detection event is recorded. The above procedure is repeated 400 times. If the percentage of detection events is larger than $95\%$, we record the $f_{\rm i}$ and $h_{\rm{s}}$ pair as a point in the sensitivity curve and move to the next frequency, otherwise we continue with a larger $h_{\rm{s}}$.

The single source GW sensitivity curves are shown in Fig.~\ref{curve}. The upper two lines are the sensitivity to GW sources with random sky-location $\theta$ and polarization $\phi$, with and without harmonic wave fitting in the timing model (thin dashed line and thin solid line respectively). The bottom two lines are the sensitivity to sources with optimal sky position and polarization with and without harmonic wave fitting in the timing model (thick dashed line and thick solid line respectively). The effect of randomization of $\theta$ and $\phi$ is to reduce the sensitivity by a factor of $\sim2.5$ from the optimal one. The harmonic wave fitting process contributes to the dents near $10^{-7}\,{\rm Hz}$. The application of FITWAVEs to remove the characteristic signal at $\sim10^{-7}\,{\rm Hz}$ affects also the nearby frequencies due to spectral leakage, but its effects are not significant, as one can see in Fig.~\ref{curve}. This is due to the fact that we have a dataset which exhibits red noise, and therefore we will have an increased detection threshold at lower frequencies. This increase in the value of the threshold smooths the effects of FITWAVE at surrounding frequencies. Spectral leakage from FITWAVEs, does not affect the estimation of the sensitivity curve at the high frequency regime in which we are interested. Each of the sensitivity curves peaks at a frequency near $1\,\rm{yr}^{-1}$ $(3.17\times10^{-8}\,\rm{Hz})$, which is caused by the pulsar position fitting process. A series of GWs with period at exactly one year would induce sinusoidal residuals in the ToAs which are indistinguishable from the effects of an offset between the real pulsar position and the initial estimated value. Therefore fitting for the pulsar position absorbs the effect of such GWs and cause loss of sensitivity at the corresponding frequency. The lost of sensitivity at low frequencies is due to the fitting of pulsar spin derivatives. At higher frequencies, the curves steadily increase with frequency with a slope of 1. The reason can be seen in equation~\eqref{amplitude}, where the GW strain of GWs needs to increase proportional to frequency in order to maintain the amplitude of the induced sinusoid in the timing residuals.

\section{Discussion}
In comparison to the sensitivity curves of PSRs J1713+0747 and J1857+0943 (B1855+09) in Y10, our sensitivity curve extends to a much higher frequency region from $4\times10^{-7}\,{\rm Hz}$ to $5\times10^{-6}\,{\rm Hz}$. Additionally, since we do not need to fit binary parameters and ``jumps" between different datasets for PSR B1937+21, our curves do not exhibit the reduced sensitivity peaks at frequencies other than $1\,{\rm yr}^{-1}$, as in Y10. The limits derived in this work are still orders of magnitude higher than that predicted by \cite{Sesana09}. To improve the limits, more precise ToAs and better DM corrections are needed.

At this point we discuss different possible causes for the unfitted noise shown in the residuals of PSR B1937+21. As shown in the middle panel of Fig.~\ref{comparesi}, the timing residuals obtained by WSRT and Jodrell Bank exhibit a synchronous trend. Therefore, we can exclude the possibility that this noise is due to clock errors at either observatory. Furthermore, the noise has no obvious frequency dependence, which indicates that it is not due to the dispersive or scattering effects of the ISM.

The remaining noise can be fitted with a sinusoidal wave of amplitude $\sim150\,\rm{ns}$ and frequency $3.4\,\rm{yr}^{-1}$ $(1.078\times10^{-7}\,\rm{Hz})$ along with its third harmonic, the amplitude of which is $\sim100\,{\rm ns}$. If this is induced by a single GW source, it means that the orbit of the SMBHB is eccentric, and the GW strain should have been at least $h_{{\rm s}}=5\times10^{-14}$ at $1.077\times10^{-7}\,{\rm Hz}$ and $h_{\rm{s}}=1\times10^{-13}$ at $3.15\times10^{-7}\,{\rm Hz}$, according to equation~\eqref{amplitude}, assuming optimal source position. These are below the sensitivity curve generated from 18 pulsars by Y10 about 0.5 dex.

However the periodic structure in the timing noise does not appear until MJD~56000. This is confirmed by splitting the dataset in two parts and computing the respective power spectra. Indeed, the power spectrum of the first part does not exhibit any distinctive peaks at low frequencies whereas the second part does. This fact argues against the possibility that this signal is due to a GW, which we would expect to be present across the whole data span.

Over a time span longer than 20 years, the noise of PSR B1937+21 is dominated by components at frequencies lower than $0.3\,\rm{year}^{-1}$. \cite{Shannon13} recently proposed that the noise structure could be explained by the existence of an asteroid belt which extends to $\sim 10\,{\rm AU}$ from the pulsar. Our study is focused on a much smaller time span, where the dominating noise components lower than $1/T_{\rm{obs}}$ are absorbed by the fitting process. Therefore a noise component with smaller amplitude at higher frequency is revealed. A planet or an asteroid in a close orbit around the pulsar with a period of 107.4 days can also create such features in the power spectrum as the upper black line in Fig.~\ref{power}. However, like as in the case of individual GWs, the structure would be expected to be seen throughout the whole data span.

Red timing noise at longer time scales has been reported \citep{Kaspi94}, and the level of which is consistent with the noise in this paper. Therefore, the most likely explanation is that this timing irregularity is intrinsic to the pulsar itself. One of the proposed mechanisms for intrinsic quasi-periodic timing noise of pulsars is a phenomenological model of the magnetic field evolution \citep{Zhang12}, in which the magnetic field of the pulsars experiences a long-term power-law decay modulated by short-term oscillations. The spin of pulsars therefore is modified by the magnetic field evolution. Moreover, some pulsars switch quasi-periodically between two different spin-down rates, as shown in \cite{Lyne10}, which could also be a possible origin for the noise structure observed in the timing residuals.

\section{Summary}
In this paper we present the timing results of PSR B1937+21 from May 2011 to May 2013, using data from the Lovell telescope, the 42-foot telescope at Jodrell Bank Observatory and the Westerbork Synthesis Radio Telescope. The  average cadence is one observation every 1.26 days, which enabled us to extend the sensitivity to GWs to a frequency $4.98\times10^{-6}\,{\rm Hz}$. Time dependent DM variations and pulse profile broadening by ISM scattering were corrected. The root-mean-square of the post-fit timing residuals of the pulsar was thus reduced to $0.346\,\rm{\mu s}$.

Unmodeled components with amplitude $\sim150\,\rm{ns}$ were observed in the post-fit residuals, inducing distinctive peaks in the power spectrum of the timing residuals at low frequencies. We found that neither clock errors at the observatories nor uncorrected dispersive ISM effects can account for this noise. If the noise was due to individual GW sources, the strain of GWs should be at least $h_{\rm{s}}=5\times10^{-14}$ at $1.077\times10^{-7}\,{\rm Hz}$ and $h_{\rm{s}}=1\times10^{-13}$ at $3.15\times10^{-7}\,{\rm Hz}$. We also discussed the possibility that this noise comes from an asteroid orbiting the pulsar with a period of 107.4 days. The transience of the noise structure make the individual GWs or asteroid origin unlikely. The most likely explanation is that the noise is intrinsic to the pulsar rotation irregularity. %Another possible origin is a small jitter in the Earth orientation, which can be confirmed or excluded with the \texttt{IRES}\footnote{http://www.iers.org/IERS/} monitoring records.

Using the noise template fitted from the power spectrum of the timing residuals, we calculated the sensitivity to individual GW sources of our data assuming optimal and random positions for the sources. The frequency regime in which we are sensitive to GWs ranges from $1.78\times10^{-8}\,{\rm Hz}$ to $4.98\times10^{-6}\,{\rm Hz}$, covering a new high frequency region never reached before by pulsar timing.

\section*{Acknowledgments}
Access to the Lovell and 42FT telescopes at Jodrell Bank is supported through an STFC consolidated grant. The Westerbork Synthesis Radio Telescope is operated by the Netherlands Institute for Radio Astronomy (ASTRON) with support from The Netherlands Foundation for Scientific Research (NWO).
SNZ acknowledges partial funding support by 973 Program of China under grant 2014CB845802, by the National Natural Science Foundation of China under grant Nos. 11133002 and 11373036, and by the Qianren start-up grant 292012312D1117210.
\bibliographystyle{mn2e}

%\bibliography{mybib}
\end{document}